\documentclass[9pt,twocolumn,twoside]{osajnl}
\journal{optica} 
\setboolean{shortarticle}{false}

\usepackage{bm}
\usepackage{commath}
\newcommand{\Ai}{\operatorname{Ai}}
\newcommand{\sgn}{\operatorname{sgn}}

\def\minititle#1{}

\title{Airy beams and accelerating waves: An overview of recent advances}

\author[1,2,7]{Nikolaos K. Efremidis}
\author[3,4,8]{Zhigang Chen}
\author[5]{Mordechai Segev}
\author[6]{Demetrios N. Christodoulides}

\affil[1]{Department of Mathematics and Applied Mathematics, University of Crete, 70013 Heraklion, Crete, Greece}
\affil[2]{Institute of Applied and Computational Mathematics, Foundation for Research and Technology - Hellas (FORTH), 70013 Heraklion, Crete, Greece.}
\affil[3]{The  MOE  Key  Laboratory  of  Weak-Light  Nonlinear  Photonics,  and  TEDA  Applied  Physics  Institute  and  School  of Physics, Nankai University, Tianjin 300457, China}
\affil[4]{Department of Physics and Astronomy, San Francisco State University, San Francisco, CA 94132}
\affil[5]{Physics Department and Solid State Institute, Technion, 3200003 Haifa, Israel}
\affil[6]{CREOL/College of Optics, University of Central Florida, Orlando, Florida 32816}
\affil[7]{nefrem@uoc.gr}
\affil[8]{zgchen@nankai.edu.cn}

\begin{abstract}
Over the last dozen of years, the area of accelerating waves has made considerable advances not only in terms of fundamentals and experimental demonstrations but also in connection to a wide range of applications.
Starting from the prototypical Airy beam that was proposed and observed in 2007, new families of accelerating waves have been identified in the paraxial and nonparaxial domains in space and/or time, with different methods developed to control at will their trajectory, amplitude, and beam width.
Accelerating optical waves exhibit a number of highly desirable attributes. They move along a curved or accelerating trajectory while being resilient to perturbations (self-healing), and, are diffraction-free.
It is because of these particular features that accelerating waves have been utilized in a variety of applications in the areas of filamentation, beam focusing, particle manipulation, biomedical imaging, plasmons, and material processing among others.
\end{abstract}

\setboolean{displaycopyright}{true}

\begin{document}

\maketitle

\section{Introduction}\label{sec:introduction}

The notion that an optical wave packet propagates along straight trajectories originates from different physical pictures such as those emerging from ray optics and electrodynamics. After all, in the absence of any index inhomogeneity, the intensity centroid of an optical beam is expected to move in a straight line - without acceleration - in full accord with Ehrenfest's theorem~\cite{messiah-2014} or the conservation of transverse electromagnetic momentum~\cite{jackson-2007}. Yet, no such restrictions exist when dealing with the  features of the intensity structure of an optical pulse or beam. Indeed, recent studies have shown that there exists a wide class of optical beams whose intensity profile does in fact transversely accelerate during propagation.  While their intensity centroid always moves in a straight line, the actual features of the intensity structure of these accelerating wavefronts follow a curved trajectory. This unusual property has significant ramifications in optics since the majority of optical processes, like nonlinear effects, detection, imaging etc., are directly dictated by the field profile itself, not by the ``center of mass'' of the beam - which is a mean property (averaged over the beam cross-section). This property holds for any system, classical and quantum, where the interaction with an external object (particles of waves) depends on the field structure of the wavepacket, not on the trajectory associated with the center of mass of the beam. Examples range from optical gradient forces acting on particles and nonlinear wave-mixing to accelerating electron-beams, where the accelerating wavepackets can propel particles on a curved trajectory, create accelerating second-harmonic beams and laser-induced plasma filaments, and even accelerate single electrons in an electron microscope. 

In section~\ref{sec:airy} we begin our review by highlighting some of the properties of Airy beams. Airy beams represent the first experimentally observed class of self-accelerating optical waves. As opposed to other diffraction-free beams previously studied, the intensity features of an Airy beam follow a curved parabolic trajectory-in a way analogous to that of a projectile moving under the action of gravity~\cite{sivil-ol2007,sivil-prl2007}.
The theory of Airy beams is developed in terms of ray or catastrophe optics, and 
three of its fundamental properties are subsequently discussed.
Interestingly, the Airy beam happens to be the only localized \textit{propagation-invariant} or \textit{diffraction-free} solution of the paraxial wave equation in one transverse dimension, and as such, its amplitude profile remains invariant in the transverse plane during propagation.
At the same time, this beam laterally shifts in the transverse plane, along a parabolic \textit{self-accelerating} trajectory-which is perhaps its most intriguing and appealing characteristic.
Finally, Airy beams exhibit \textit{self-healing} properties, in the sense that they tend to reform or reconstitute themselves even when they have been severely impaired or perturbed. 
Other classes of solutions having an Airy profile that can be synthesized in space and time, with separable or correlated profiles, are also presented.

In sections~\ref{sec:paraxial}-\ref{sec:bessel-like} we discuss various families of accelerating waves that are possible in paraxial as well as non-paraxial settings where a non-paraxial acceleration can lead to large deflection angles.
Note that in the nonparaxial regime we have to redefine the notion of diffraction-free (or propagation-invariant or non-diffracting) so that its amplitude profile remain invariant in the plane that is transverse to the trajectory. 
In analyzing accelerating waves, two different approaches have been developed.
The first one is to utilize different coordinate systems in order to find analytic expressions either in closed-form or in integral form.
The other is to use a generic ray optics/catastrophe theory methodology to derive accelerating beams with pre-designed properties (trajectory, amplitude, and beam width). 

In section~\ref{sec:aaf} we focus on the properties of radially symmetric accelerating beams or abruptly autofocusing waves. Such optical waves have a broad intensity profile up until the focal plane where the entire beam focuses abruptly, and consequently its peak intensity abruptly increases within a small focal volume. We discuss different classes of abruptly autofocusing beams and provide a theoretical description of their focusing properties such as their focal distance, profile, and contrast. 

The nonlinear behavior of accelerating beams is discussed in section~\ref{sec:nonlinear}. We analyze the dynamics of Airy beams in nonlinear environments, and we present families of nonlinear accelerating waves and their instabilities and dynamics in Kerr, quadratic, and saturable nonlinear media.

In section~\ref{sec:nonuniform} we consider the dynamics of Airy and other accelerating waves in media with non-uniform index distributions such as for example in linear index potentials that can be utilized to dynamically control the trajectory of the beam. In addition we consider the propagation of accelerating beams in parabolic and other types of potentials such as periodic media, as well as their reflection and refraction from interfaces.

Applications of Airy, accelerating, and abruptly autofocusing waves in different fields are discussed in section~\ref{sec:applications}. Areas such as filamentation, biomedical imaging, particle manipulation, material processing, and plasmons greatly benefit from the properties of these accelerating optical waves. 

Finally, in section~\ref{sec:settings} we elaborate on other settings where accelerating waves have been suggested and observed such as electron beams and relativistic fermion beams, water waves, sound waves, curved space, and Bose-Einstein condensates.

Over the past decade, interest in accelerating waves has surged. Of course, as with any other review on this topic~\cite{hu-springer2012,bandr-opn2013,minov-lpr2014,levy-po2016}, this article has its own focus and is by no means complete.

\section{Airy waves}\label{sec:airy}

\begin{figure}
\centering
\includegraphics[width=\linewidth]{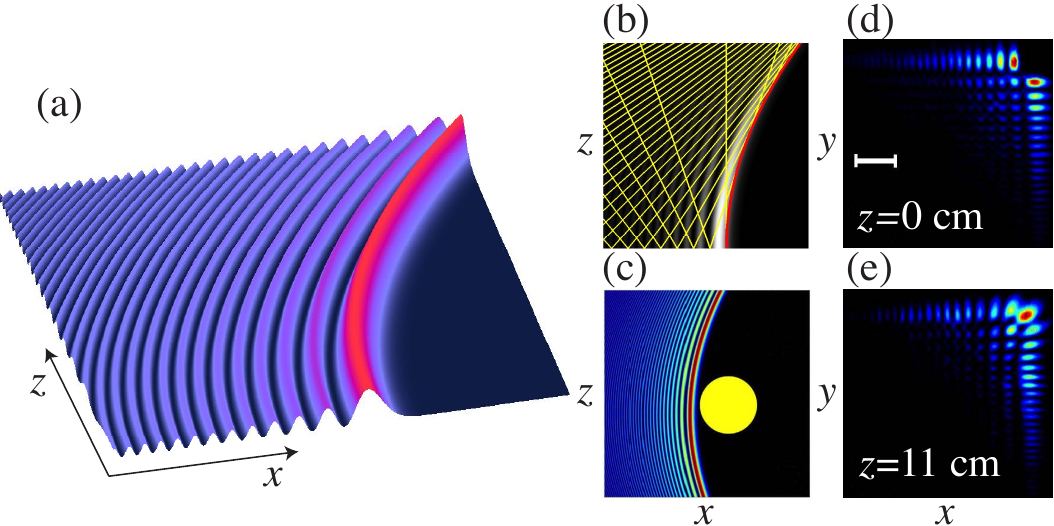}
\caption{An Airy beam along with its nondiffracting, self-bending, and self-healing properties. In (a) the propagation dynamics of an accelerating diffraction-free Airy beam is depicted. A ray optics description is shown in (b) with the rays presented as yellow lines and the parabolic caustic trajectory shown with a red curve. (c) Due to its curved parabolic trajectory, the Airy beam is able to circumvent obstacles.
  The experimental generation of a two-dimensional Airy beam is shown in (d)-(e).
  (d) On the input plane the main lobe of the Airy beam is blocked. (e) Due to self-healing the main lobe is then reconstructed during propagation. The scaling bar in (d)-(e) is 200 $\mu$m.}
\label{fig:airy}
\end{figure}

\minititle{Airy beam solution}

The concept of self-accelerating wavepackets emerged in 1979 with the pioneering work of Berry and Balazs~\cite{berry-ajp1979}, who found a shape-preserving accelerating solution to the potential-free Schr\"odinger equation in the form of an ideal Airy function.
The term \textit{self-accelerating} indicates that the wave is accelerating without the need of an external potential. 
Remarkably, this result remained relatively unnoticed for decades. To some extent this may be due to the fact that is difficult to experimentally prepare quantum particles in an Airy state or because the Airy wave itself has an infinite norm.        

The study of optical accelerating waves effectively begun with the understanding that what really matters is not the trajectory of the ``center of mass'' but the curved evolution of the field itself. Thus, a truncated Airy beam (which is physically realizable since the beam must have finite power), interacts with its surroundings - whether they are particles or other waves - according to its accelerating field structure, not according to its centroid which moves instead on a straight line so as to ensure the preservation of transverse momentum. With this understanding in mind, the first prediction and observation of Airy beams took place in 2007~\cite{sivil-ol2007,sivil-prl2007}, almost 30 years after the Berry and Balazs pioneering work.

Specifically, it was shown~\cite{sivil-ol2007,sivil-prl2007} that, in one transverse dimension, the paraxial wave equation 
\begin{equation}
i\psi_z+\frac1{2k}\psi_{xx}=0
\end{equation}
supports an Airy wave solution of the form
\begin{equation}
\psi=\Ai\left(\gamma x-\frac{\gamma^4z^2}{4k^2}+i\frac{\alpha\gamma z}{k}\right)\\
e^{i\phi(x,z)+\alpha\left(x-\frac{\gamma^3z^2}{2k^2}\right)}
\label{eq:airy_sol}
\end{equation}
where $x$ is the transverse and $z$ is the longitudinal coordinate, $k$ is the wavenumber, $\gamma$ is inversely proportional to the pulse width, and $\phi=(\gamma^2z/(2k))(\gamma x-\gamma^4z^2/(6k^2)+\alpha^2/\gamma^2)$ is the accumulated phase during propagation. The parameter $\alpha$ determines the degree of apodization or truncation of the Airy beam which keeps its total power $P=e^{(2/3)(\alpha/\gamma)^3}/\sqrt{8\pi\alpha\gamma}$ finite. The Fourier transform of the Airy wave solution is given by a Gaussian beam with an additional cubic phase 
\begin{equation}
\tilde\psi(k) = (1/\gamma) e^{i(k+i\alpha)^3/(3\gamma^3)}.
\end{equation}
Thus, it can be generated by simply imposing a cubic phase to a Gaussian laser mode, for example, by utilizing a spatial light modulator. By setting the truncation factor to zero ($\alpha=0$) in Eq.~(\ref{eq:airy_sol}) the ``ideal'' infinite-power Airy beam is recovered~\cite{berry-ajp1979}. 
As previously mentioned, this corresponds to the Airy wavepacket first obtained within the context  of quantum mechanics by solving the potential-free Sch\"odinger equation - an equation that is mathematically equivalent to the paraxial equation of diffraction.

\minititle{Diffraction-free + parabolic trajectory} 
The Airy beam exhibits several unique properties which we will discuss in this section.  Since there are different methods for truncating an Airy wave we are going to focus on the ideal form of the solution [Eq.~(\ref{eq:airy_sol}) with $\alpha=0$]. In Fig.~\ref{fig:airy}(a) we depict the intensity dynamics of an Airy beam. We observe that the solution remains propagation-invariant (or diffraction-free) in the transverse plane for an observer that follows the parabolic trajectory $x=\gamma^3z^2/(4k^2)$. Actually, the Airy wave has been formally proven to be the only diffraction-free solution of the paraxial equation in the case of one transverse dimension~\cite{unnik-ajp1996}. That is, in one transverse dimension there is no shape-preserving wavepacket that is evolving on a straight line (zero acceleration), unlike the case of two transverse dimensions which where entire families of on-axis propagation-invariant solutions exist (e.g. Bessel, Mathieu beams, etc.)   

\minititle{Theory} 
Theoretical analysis of Airy beams (and more generally of accelerating beams)
can be carried out by utilizing fundamental principles of ray optics and catastrophe theory~\cite{berry-1980,kravt-1999}. The starting point in the case of one transverse dimension, is selected to be the Rayleigh-Sommerfeld diffraction formula~\cite{penci-ol2015}
\begin{equation}
\psi(x,z) = 2\int_{-\infty}^\infty \psi_0(\xi)
\frac{\partial G(x,z;\xi)}{\partial z}\dif\xi. 
\label{eq:green}
\end{equation}
In Eq.~(\ref{eq:green}) $\psi_0(x)=A(x)e^{i\phi(x)}$ is the amplitude and phase decomposition of the optical wave on the input plane, $\xi$ is an ``$x$-conjugated'' transverse coordinate on the input plane, $k=n\omega/c=2\pi/\lambda$, $G(x,z;\xi)= -(i/4)H_0^{(1)}(kR)$, $H_0^{(1)}$ is a Hankel function, and $R=((x-\xi)^2+z^2)^{1/2}$.
Asymptotically the phase of the integrand is given by $\Psi=kR+\phi(\xi)-\pi/4$. 
Under the paraxial approximation $z\gg|x-\xi|$, and after utilizing the Airy beam as initial condition, the resulting ray equation
\begin{equation}
x=\xi\pm\gamma (-\gamma\xi)^{1/2}z/k
\label{eq:airy:rays}
\end{equation}
is obtained by setting $\partial_\xi\Psi=0$. The additional condition $\Psi_{\xi\xi}=0$ for an optical catastrophe determines the high intensity parabolic trajectory (or caustic)
\begin{equation}
x_c=\gamma^3z_c^2/(4k^2)
\label{eq:airy:cat}
\end{equation}
of the Airy solution, where in the above formula the subscript $c$ stands for the caustic. The rays of the Airy beam are depicted with yellow lines and the caustic with a red curve in Fig.~\ref{fig:airy}(b). We clearly see that the caustic separates the $x-z$ plane into two regions with different characteristics: On the left side two rays intersect at each point whereas on the right side no rays exist. As we approach the caustic from the left side the intersecting rays tend to become almost parallel to each other, a geometric criterion that signifies the presence of a caustic. 

\minititle{Balistic control} 
The parabolic trajectory of an Airy beam can be further engineered. As it was experimentally demonstrated in~\cite{sivil-ol2008}, by applying an additional linear phase on the input plane, the Airy beam can now follow a predetermined ballistic motion. Importantly, this property can be utilized to generate beams that can circumvent an obstacle [Fig~\ref{fig:airy}(c)] - a functionality the could be useful, for example, in particle manipulation. Furthermore, the projectile motion of two-dimensional truncated Airy beams along general ballistic trajectories with controllable range and height was also demonstrated~\cite{hu-ol2010}. 

\minititle{Self-healing} Another fundamental property of the Airy beam is its ability to self-reconstruct or self-heal during propagation~\cite{broky-oe2008}. For example, in Fig.~\ref{fig:airy}(d)-(e) we see the experimental results pertaining to a 2D Airy beam, which is constructed by separation of variables from two 1D Airy beams. In this case, the main lobe (or the corner lobe in the 2D case) is initially obstructed. We observe that this lobe as well as the whole 2D Airy beam is then reconstructed during propagation. 
The principle of operation behind this ``self-healing'' mechanism can be understood in terms of wave superposition based on Babinet's principle~\cite{bouch-oc1998}. If $v(\bm r_\perp,z=0)$ is an arbitrary localized ``obstruction'' of any propagation-invariant solution $\psi(\bm r_\perp,z=0)$ (such as the Airy beam)  on the input plane $u=\phi+v$, then the obstruction itself is subjected to diffraction. Thus, as $z$ increases, $v\rightarrow0$  and the diffraction-free beam is reconstructed as shown in Fig~\ref{fig:airy}(e). The self-healing property is very important for optical waves propagating in turbulent or other adverse environments. The robustness of an Airy beam propagating in scattering and turbulent media was examined in~\cite{broky-oe2008}. It was demonstrated experimentally that due to its diffraction-free and self-healing properties, the Airy beam can retain its shape under turbulent conditions where a Gaussian beam would suffer massive deformations in the same environment. The robustness of the Airy beam has also been confirmed in different kind of turbulent settings~\cite{hu-ol2010,gu-ol2010,chu-ol2011}. 

\begin{figure}
\centerline{\includegraphics[width=\columnwidth]{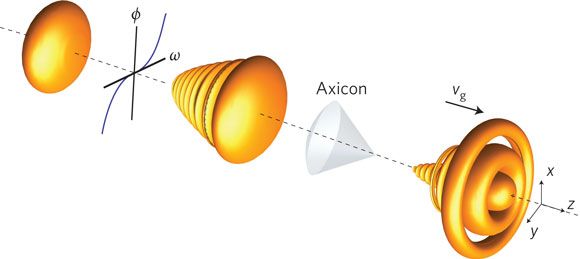}}
\caption{Experimental generation of an Airy-Bessel light bullet that propagates undistorted in a three-dimensional space-time environment. The temporal Airy profile is generated by adding a large cubic spectral phase whereas the spatial Bessel profile is produced by an axicon.
  Reprinted with permission from Springer Nature~\cite{chong-np2010}.}
\label{fig:bullet}
\end{figure}
\minititle{Light bullets} 
Due to the mathematical similarity between the effect of paraxial diffraction in space and second-order dispersion in time, one can show that dispersion-free Airy pulses with accelerating space-time characteristics are also possible in dispersive systems. In this respect, by employing separation of variables, it is possible to obtain a three-dimensional field configuration that is localized in space-time and as such it remains impervious to the combined effects of dispersion and diffraction. The first proposal for such an optical bullet was based on the compensation of diffraction and anomalous dispersion by the self-focusing Kerr nonlinearity~\cite{silbe-ol1990}. However, such an structure is inherently unstable due to the effect of catastrophic wave collapse. However, engineering a spatiotemporal Bessel-Airy (or triple Airy) wavepacket in a simple linear dispersive medium (such as glass) can circumvent this problem and give rise to a light bullet~\cite{sivil-ol2007}. Indeed, by combining the diffraction and dispersion-free profiles of a Bessel beam~\cite{durni-prl1987,durni-josaa1987} and a temporal Airy pulse (Fig.~\ref{fig:bullet}), a three-dimensional linear localized light bullet was observed~\cite{chong-np2010}. Subsequently, using a different approach, an optical bullet comprised of three Airy wavepackets in the two spatial transverse coordinates as well as in time was demonstrated~\cite{abdol-prl2010}. Different families of optical bullets can be generated either as oblique Airy waves~\cite{eiche-ol2010} or by using separation of variables~\cite{zhong-pra2013,deng-oe2016}. 

\minititle{Airy pulses} 
Higher-order dispersion can play an important role, especially in the dynamics of ultrashort Airy pulses. The dispersion dynamics of a Gaussian pulse under the action of cubic dispersion in a single-mode fiber, were first addressed in~\cite{miyag-ao1979-1} where it was found that asymptotically the pulse develops an Airy-type profile. The problem associated with the propagation of Airy pulses in optical fibers, under the influence of third-order dispersion has been analytically solved in~\cite{besier-pre2008}. In this study, it was shown that the Airy pulse distortion caused by third order dispersion is not all that significant. On the other hand, close to the zero-dispersion point of the fiber and for particular choices for the second and third order dispersion, the Airy pulses can undergo focusing followed by the inversion of their temporal profile~\cite{dribe-ol2013}. 

\minititle{Correlated space-time profile} 
Very recently, classes of diffraction-free pulsed beams with a correlated profiles in space-time have attracted research attention~\cite{konda-oe2016,parke-oe2016}. 
In the case of a parabolic conic section of the dispersion cone and a cubic phase excitation in the Fourier space the resulting wavepacket takes the form of a non-paraxial Airy pulsed-beam~\cite{efrem-ol2017-fsw}. Such optical waves have a self-bending parabolic spatial profile and can propagate in the forward or in the backward direction with luminar speeds.  A paraxial class of Airy beams with correlated space-time profiles has been observed in~\cite{konda-prl2018}. The acceleration-free Airy wave packet that travels in a straight line was synthesized by deforming its spatiotemporal spectrum to reproduce the effect of a Lorentz boost.

Beams with curved or accelerating profile are not limited to solutions of the Airy type. A significant amount of progress had been made in identifying different classes of beams with other unique properties as well as in generating accelerating beams with pre-engineered trajectories, amplitudes, and beam widths.  

\section{Other classes of paraxial accelerating waves}\label{sec:paraxial}
\minititle{Paraxial 2D solutions}
In the paraxial regime and in the case of one transverse dimension, it has been formally proven that the only diffraction-free solution happens to be the Airy beam itself. However, in the case of two transverse dimensions, apart from the 2D Airy beams~\cite{sivil-prl2007,sivil-ol2008,hu-ol2010}, different classes of exact solutions have also been found by following an operator splitting approach~\cite{bandr-ol2008}. The solutions were constructed in an orthogonal parabolic basis, giving rise to a Schr\"odinger-type equation with a quartic potential. Subsequently such beams were experimentally observed~\cite{davis-oe2008}. Additional classes of parabolic propagation-invariant accelerating beams were also found using Fourier transforms~\cite{bandr-ol2009}. 

\minititle{Incoherent}
The dynamics of partially coherent Airy beams that pass through the same cubic phase mask has been studied experimentally~\cite{morri-oe2009} and theoretically~\cite{dong-olt2013,eyyub-apb2013}. In a different approach, by superimposing accelerating beams in the paraxial and the nonparaxial domains that are associated with the same curved paths, it was found to be possible to generate incoherent self-accelerating beams with well-defined trajectories~\cite{lumer-optica2015}. The non-paraxial version of such partially-coherent accelerating beams is most interesting, as they are superpositions of multiple high-order solutions of the non-paraxial Helmoltz equation, all propagating on the same curved trajectory~\cite{lumer-optica2015}.

\minititle{Enhanced Airy beams}
The question whether it is possible to generate optical waves with enhanced properties has been examined in different works by mainly utilizing the spectral properties of the accelerating beams. In~\cite{bongi-sr2015} two-dimensional accelerating beams following different trajectories were demonstrated by using a Fourier space generation method. It is shown that an appropriate transverse compression of their spectra leads to a dramatic enhancement of their peak intensity and a significant decrease of their transverse expansion. A similar approach was applied in the case of spatiotemporal Airy bullets~\cite{bongi-oe2016}. The resulting waves show significant enhancement of the central lobe intensity while having reduced spatiotemporal outspread. Using a different approach, by multiplying the two-dimensional Fourier spectrum cubic phase with a binary function, results to the generation of a ``super-Airy'' beam with a parabolic trajectory and a main lobe that is reduced to nearly half the size and increases intensity as compared to the Airy beam~\cite{singh-ol2015}.

\minititle{Paraxial catastrophes}
In terms of catastrophe theory, the Airy beam is described as a fold catastrophe~\cite{zeema-1977}. This is the simplest possible catastrophe with one internal and one control variable. The dynamics of higher order catastrophes have also been explored theoretically and experimentally. In particular catastrophes of the Pearcey type were shown to exhibit autofocusing properties~\cite{ring-oe2012}. By utilizing catastrophe theory, different classes of beams can be synthesized in the paraxial domain. These include Pearcey, umbilic and swallowtail catastrophes~\cite{berry-jo2017}. Furthermore, optical beams that are governed by a swallowtail catastrophe were observed in~\cite{zanno-optica2017}.
\minititle{Superposition of Gaussian beams}
Using a different approach that does not rely on the use of caustics or of an extended focus, paraxial shaping of light in three-dimensions is presented by superimposing multiple Gaussian beams along an arbitrary curve~\cite{rodri-oe2013}.

\begin{figure}
\centerline{\includegraphics[width=\columnwidth]{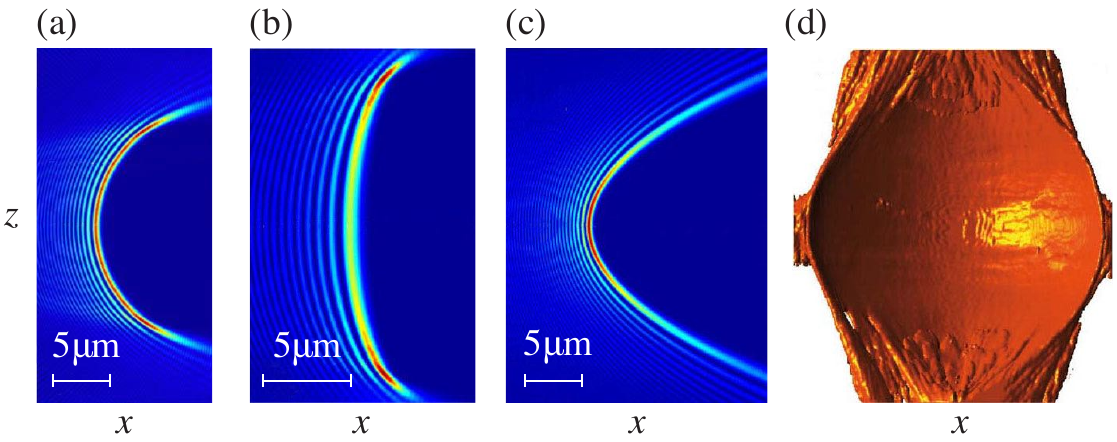}}
\caption{Different classes of nonparaxial accelerating beams. (a) Bessel-type diffraction-free accelerating beam that follows a circular trajectory. (b) Mathieu and (c) Weber accelerating beams. (d) Experimental results for a spherically shaped field.
}
\label{fig:nonparaxial}
\end{figure}
\section{Nonparaxial accelerating waves}\label{sec:nonparaxial}
\minititle{Nonparaxial}
The families of solutions discussed up until this point are paraxial and thus can only bend at relatively small angles. This was verified in the case of an Airy beam which was used as an initial condition to Maxwell's equations~\cite{novit-ol2009}. In the non-paraxial domain, by following a ray optics approach, arbitrary micro-scale caustic beams were constructed in the case of two and three dimensions~\cite{froeh-oe2011}. The resulting waves can bend up to a $60^\circ$ angle. In addition, light sheets were produced by applying an additional rotational phase. The problem that was examined in~\cite{kamin-prl2012} is whether there exists a nonparaxial generalization of the Airy beam with diffraction-free properties. Starting from the Helmholtz equation, it was found that diffraction-free non-paraxial solutions of the Bessel-type do exist-satisfying Maxwell's equations.  Specifically, the Helmholtz equation
\begin{equation}
  E_{xx}+E_{zz}+k^2E=0
\end{equation}
where $x$ is the transverse coordinate, $z$ is the propagation coordinate, and $k=\omega/c$ is the wavenumber in the homogeneous linear medium, supports Bessel solutions of the form
\begin{equation}
  E = J_\alpha(kr)e^{i\alpha\theta}
\end{equation}
where $r$ and $\theta$ are the polar coordinates $x=r\cos\theta$, $z=r\sin\theta$. 
Since Bessel beams are synthesized from both forward and backward propagating plane waves, a selection was used to keep the forward and discard the backward propagating modes. The resulting ``half Bessel'' beam takes the form~\cite{kamin-prl2012}
\begin{equation}
  J_\alpha^+(kx,kz) = \frac1{2\pi}\int_0^\pi
  e^{i\alpha k_\theta+ik[x\cos k_\theta+z\sin k_\theta]}\,\dif k_\theta. 
\end{equation}
These solutions follow a circular trajectory, in contradistinction from the Airy which follow a parabolic path. The radius of the circular acceleration depends on the value of $\alpha$. A typical example of such a beam propagating in a shape-preserving form on a circular trajectory bending at almost $180^\circ$ is shown in Fig.~\ref{fig:nonparaxial}(a). Nonparaxial diffraction-free beams following circular trajectories were subsequently experimentally observed in~\cite{courv-ol2012,zhang-ol2012}. Shortly thereafter, new classes of exact accelerating solutions of the Helmholtz equation were found, having the form of Mathieu and Weber functions exhibiting shape-preserving propagation on elliptic and parabolic trajectories~\cite{zhang-prl2012,aleah-prl2012,bandr-njp2013} [Figs.~\ref{fig:nonparaxial}(b)-(c)]. Note that, mathematically speaking, these solutions cannot be considered as ideally diffraction-free since their amplitude profile does not remain invariant in the plane that is normal to the trajectory.
However, physically, this limiting argument is common also to ideal Bessel beams and Airy beams, which should carry infinite power, and therefore physically must be truncated - which always sets a fundamental limit on their range of propagation in a shape-preserving form. One important difference between the paraxial Airy beams and the non-paraxial accelerating beams is that the former is a unique solution (i.e., there is only one Airy beam associated with a given parabolic trajectory), whereas the latter admits a whole family of solutions, all propagating on the same circular (or elliptic) trajectory.
Interestingly, using symmetry arguments, nonparaxial accelerating beams can create high intensity surfaces such as spherical and spheroidal~\cite{aleah-prl2012,alons-ol2012}. Spherical shaping of light has been experimentally demonstrated in~\cite{mathi-ol2013} as shown in Fig.~\ref{fig:nonparaxial}(d). An efficient method to generate nonparaxial accelerating beams is to utilize the Fourier space properties of these solutions~\cite{hu-pra2013,mathi-ol2013}. In this case, exact formulas have been obtained connecting the phase in the Fourier space with the caustic trajectory. Accelerating beam can be generated by the use of mirrors~\cite{alons-oe2014-1,alons-oe2014-2}. In this case it is possible to generate caustics that extend well beyond the semicircle. 

\begin{figure}[t]
\centerline{\includegraphics[width=\columnwidth]{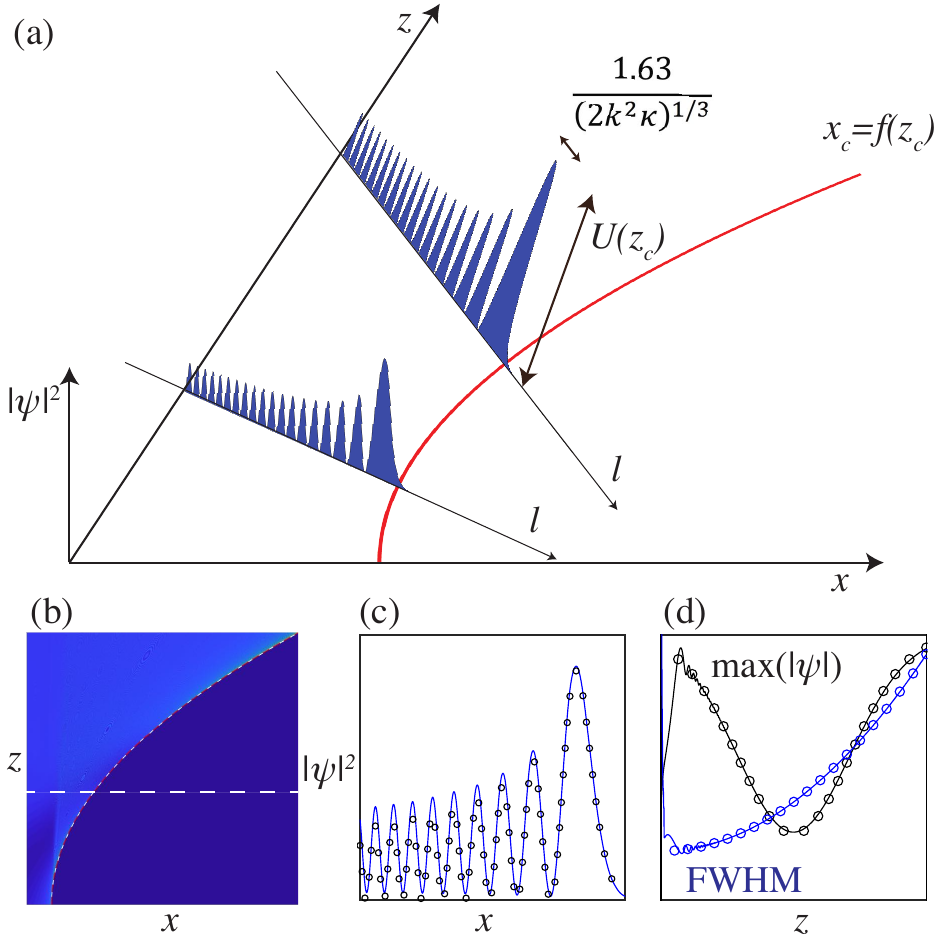}}
\caption{(a) Accelerating beams with a pre-designed trajectory $x_c=f(z_c)$ and maximum amplitude $U(z_c)$ along the path. The beam width depends on the curvature of the trajectory $\kappa$ and the wavelength. (b)-(d) An example of a nonparaxial beam that follows a parabolic trajectory with amplitude profile $U(z_c)=A-Be^{(z_c/z_0)^2}$. (b) Amplitude dynamics (c) intensity cross section [corresponding to the dashed line in (b)], (d) maximum amplitude and full width at half maximum as a function of propagation distance $z$. In (c)-(d) theoretical (circles) and numerical (solid curves) results are compared.
}
\label{fig:accelerating}
\end{figure}
\minititle{Design of the trajectory}
An important question is whether accelerating beams can be synthesized to follow arbitrary functional forms of the trajectory $x_c=f(z_c)$. 
The use of caustics imposes an inherent limitation to the geometry of the trajectory which has to be convex. Besides this limitation, the answer to the question is positive, and in fact it been examined in parallel in three different publications~\cite{green-prl2011,chrem-ol2011,froeh-oe2011}. The resulting expression between the derivative of the phase on the input plane and the pre-designed trajectory in the nonparaxial regime is found to be given by~\cite{froeh-oe2011}
\begin{equation}
\frac{\dif\phi(\xi)}{\dif\xi}=
k\frac{\dif f/\dif z}{[1+(\dif f/\dif z)^2]^{1/2}}.
\label{eq:tangent}
\end{equation}
Equation~(\ref{eq:tangent}) can, in general, be integrated numerically and provides the required phase information for a predefined convex functional form of the beam path. However, there are particular classes of trajectories including circular, elliptic, power-law, and exponential trajectories whose phase profile is given in closed-form  both in the paraxial~\cite{green-prl2011,chrem-ol2011,froeh-oe2011} and in the nonparaxial regime~\cite{penci-ol2015}. 

\minititle{What is diffraction-free}
In the paraxial regime a localized beam is defined as propagation-invariant or diffraction-free if its amplitude profile remains invariant in the transverse plane as it propagates. In the non-paraxial regime accelerating beams can bend at large angles and thus the notion as to what is diffraction-free must be accordingly modified. Specifically, we can define a beam to be propagation-invariant or diffraction-free if its amplitude profile remains invariant in the plane that is normal to the trajectory [see Fig.~\ref{fig:accelerating}(a)]. Since the most important feature of a diffraction-free beam is its main lobe, such a calculation can even be local (valid close to the beam trajectory). 

\minititle{Engineering the trajectory, the amplitude, and the beam width}
Most of the works in the area of accelerating beams focused on engineering the trajectory of the beam and did not take into account additional degrees of freedom. 
As it was recently shown, it is possible to generate paraxial and non-paraxial optical beams with pre-engineered trajectories and pre-designed maximum amplitude~\cite{gouts-pra2018,gouts-oe2018}. The independent control of these two degrees of freedom is made possible by engineering both the amplitude and the phase of the optical wave on the input plane. 
Since the beam width depends solely on the local curvature of the trajectory, it is also possible to generate beams with pre-defined amplitudes and beam-widths. 
From Eq~(\ref{eq:green}) the amplitude profile of an accelerating beam close to the caustic is given by~\cite{gouts-oe2018}
\begin{equation}
\psi = 
2A(\xi)
\left(\frac{\pi^4 R_c^3\kappa^2}{\lambda}\right)^{1/6}
e^{i\Xi}
\Ai\left(-s[2k^2\kappa(z_c)]^{1/3}\delta l\right)
\label{eq:psi:asy}
\end{equation}
where $s=\sgn(z_c'(\xi))$, $\kappa$ is the curvature of the trajectory
\begin{equation}
\kappa(z_c) = 
\frac{\left|f''(z_c)\right|}
{[1+(f'(z_c))^2]^{3/2}}
=\frac{z_c^2}{R_c^3|z_c'(\xi)|},
\end{equation}
$\Xi=kR_c+\phi-\pi/4$ is an additional phase factor, $R=\sqrt{(x-\xi)^2+z^2}$, $R_c=R(x_c,z_c,\xi)$ and for simplicity we have replaced $\xi_c$ with $\xi$. In the above expression $l$ is a local coordinate that is perpendicular to the trajectory [see Fig.~\ref{fig:accelerating}(a)] and $\delta l=0$ exactly at the caustic $(x_c,z_c)$. A similar formula is obtained by working in Cartesian coordinates $(x,z)$ and using the relation $\delta l=z_c(\delta x)/R_c$ in the argument of the Airy function. From Eq.~(\ref{eq:psi:asy}) the inverse problem of generating accelerating beams with a pre-designed trajectory (or beam width) and designed maximum amplitude along the trajectory $U(z_c)$ can be solved.
The required amplitude on the input plane is then given by
\begin{equation}
A(\xi)
=
\frac{U(\xi)}{2.3}
\left(\frac{\lambda}{R_c^3 \kappa^2(z_c)}\right)^{1/6}. 
\label{eq:ampl}
\end{equation}
In addition, from the argument inside the Airy function in Eq.~(\ref{eq:psi:asy}), we can determine the beam full width at half maximum
\begin{equation}
W(\xi) \approx \frac{1.63}{[2k^2\kappa(z_c)]^{1/3}}
\label{eq:bw}
\end{equation}
which, besides the wavelength, depends only on the curvature of the trajectory. 
A typical example of a nonparaxial accelerating beam with pre-designed trajectory and  amplitude is shown in Fig.~\ref{fig:accelerating}.
The paraxial limit of Eq.~(\ref{eq:psi:asy}) can be obtained by setting $\kappa=|f''(z_c)|=1/[z_cz_c'(\xi)]$ and 
assuming $z_c\gg|x_c-\xi|$. 
A direct outcome of the asymptotic calculations is that only beams with constant curvature can have constant beam width and thus are potentially diffraction-free. In one transverse direction, in the nonparaxial domain, the only trajectory with constant curvature is the circular (Bessel beams~\cite{kamin-prl2012,courv-ol2012,zhang-ol2012}), and in the paraxial is the parabolic (Airy beams~\cite{sivil-ol2007,sivil-prl2007}). 

\begin{figure}
\centerline{\includegraphics[width=\columnwidth]{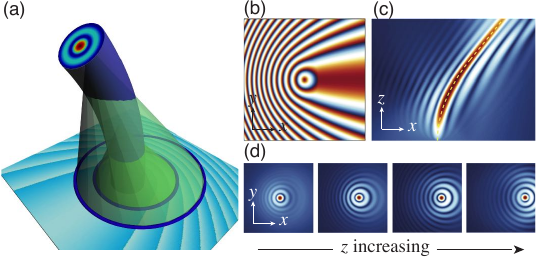}}
\caption{Bessel-like beams. (a) Rays emitted from expanding circles on the input plane intersect along the focal curve. (b) Phase profile on the input plane, (c) intensity cross section along the $y=0$ plane, and (d) cross sections on the transverse plane for increasing $z$.
}
\label{fig:Bessel}
\end{figure}

\section{Bessel-like beams}\label{sec:bessel-like}
\minititle{Bessel-like}
All the families of solutions discussed so far rely on the same physical principle (i.e., the presence of caustics) for their generation. 
Note however, that the Bessel beam~\cite{durni-prl1987,durni-josaa1987}, a member of another class of diffraction-free beams, propagates along straight trajectories and is generated by a different physical mechanism: 
rays emitted from expanding circles on the input plane intersect along a focal line. In subsequent works it was found that it is possible to generate Bessel-like beams that propagate along spiraling~\cite{jarut-ol2009,matij-oe2010,morri-jo2010} and snake or zigzag~\cite{rosen-ol1995} trajectories. 
A general methodology for generating diffraction-free Bessel-like beams that propagate along arbitrary trajectories was proposed in~\cite{chrem-ol2012-bessel} and subsequently observed in~\cite{zhao-ol2013}. The method relies on finding a continuous focal curve which is the locus of the apexes of conical ray bundles emitted from expanding and shifting circles on the input plane as shown in Fig.~\ref{fig:Bessel}(a). An advantage of the Bessel-like beams in comparison to caustic-based beams is that the former can follow trajectories that are not necessarily convex. A typical example of a Bessel-like beam is shown in Fig.~\ref{fig:Bessel}(b)-(d). Extending the analysis in the nonparaxial regime gives rise to nonparaxial Bessel beams that follow pre-defined trajectories with large bending angles and subwavelength features~\cite{chrem-pra2013}. 
The same principle of operation can be applied to generate higher-order vortex Bessel-like beams~\cite{zhao-sr2015}. In this case, rays emitted from expanding circles on the input plane at skew angles with respect to the propagation axis, interfere to create an oblique cone-like surface with a constant minimum waist.
Such a judiciously shaped singular beam can be navigated along an arbitrary yet pre-determined trajectory while preserving its angular momentum and its associated dark ``hole'' within the main lobe of the beam~\cite{zhao-sr2015}.

\begin{figure}
\centerline{\includegraphics[width=\columnwidth]{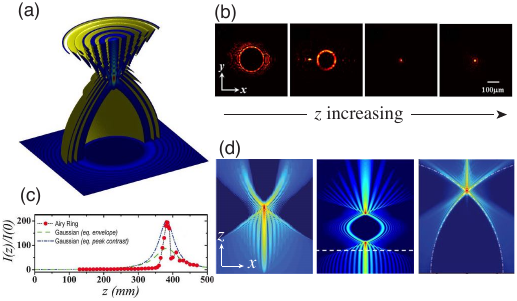}}
\caption{Abruptly autofocusing beams. (a) 3D plot of the intensity dynamics. (b) experimental results. (c) Comparison of the intensity contrast between comparable abruptly autofocusing beams and Gaussian beams. (d) Side views of an abruptly autodefocusing beam, a bottle beam, and a nonparaxial autofocusing beam (from left to right).
}
\label{fig:aaf}
\end{figure}

\section{Abruptly autofocusing waves}\label{sec:aaf}

\minititle{Autofocusing beams}
The focusing characteristics of optical beams have always been an issue of great practical importance. In the case of a Gaussian wavefront - perhaps the most prevalent of all beams - the peak intensity follows a Lorentzian distribution around the focus  (minimum waist point). For many applications it is crucial that a beam abruptly focuses its energy right before a target while maintaining a low intensity profile until that very moment. 
In several experimental settings, such a behavior can be useful in suddenly ``igniting'' a particular nonlinear process, such as multiphoton absorption, stimulated Raman, and optical filaments in gases, where it is desirable to begin the process locally, only after the focal plane. 
In order to realize this fascinating prospect, a class of accelerating waves was proposed~\cite{efrem-ol2010} with an intensity peak that remains almost constant during propagation, while close to a particular focal point it suddenly autofocuses and, as a result, the peak intensity can abruptly increase by orders of magnitude. The first family of abruptly autofocusing waves suggested was generated from a radially symmetric Airy beam of the form $\Ai(r_0-r)\exp[\alpha(r_0-r)]$
where $r_0$ is the radius of the Airy ring on the input plane and $\alpha$ is an exponential truncation factor. Such waves were experimentally observed in~\cite{papaz-ol2011,zhang-ol2011}. The dynamics of the Airy ring are shown in Fig.~\ref{fig:aaf}(a) (theory) and Fig.~\ref{fig:aaf}(b) (experiment). 
A comparison of the focusing characteristics between comparable autofocusing and Gaussian beams illustrates the sharper contrast and higher peak intensity associated with the Airy ring~\cite{papaz-ol2011}, as shown in Fig.~\ref{fig:aaf}(c). 

\minititle{Asymptoptics}
By engineering the phase profile on the input plane, classes of abruptly autofocusing waves that follow different trajectories were proposed~\cite{chrem-ol2011}. Such beams can be efficiently generated in the Fourier space~\cite{chrem-ol2011-fourier,chrem-pra2012} by phase modulating a circularly symmetric optical wavefront, such as a Gaussian, and subsequently applying a Fourier transform by a means of a lens. For an autofocusing beam that follows an arbitrary caustic trajectory $r_c=f(z_c)$, asymptotic expressions for the amplitude dynamics were obtained in~\cite{gouts-pra2018}. Using these formulas, expressions for the trajectory, the intensity contrast, and the beam width at the focus were derived in terms of the geometric properties of the trajectory~\cite{gouts-pra2018}. Specifically, it is found that the location of the focus $z_f$ is shifted forward as compared to the location of the caustic
\begin{equation}
z_f = z_c -\frac{1}{(2k^2\kappa(z_c))^{1/3}g(z_c)}
\label{eq:zf}
\end{equation}
where $z_c=z_c(r_c=0)$, $\kappa=|f''(z_c)|$ is the paraxial curvature, and $g=f'(z_c)<0$. In addition, the intensity contrast at the focus 
\begin{equation}
\frac{I_{\mathrm{max}}(z=z_f)}{I_{\mathrm{max}}(z=0)}
\approx
17.98
\left(
\frac{(k^2\kappa)^{1/3}\rho_c A(\rho_c)}{\max(A(\rho_c))}
\right)^2
\end{equation}
depends on the amplitude $A(\rho_c)$, the radial distance from the axis on the input plane $\rho_c$ of the ray that converges at the focus, and the curvature $\kappa(z_f)$. Finally the amplitude profile on the focal plane is given by 
\begin{multline}
\psi(r,z)=
\sum_{j=1,2}
\rho_{c,j}A(\rho_{c,j})
\left|\frac{2\pi kz_{c,j}} {z(z_{c,j}-z)}\right|^{1/2}
J_0\left(\frac{kr\rho_{c,j}}{z}\right) 
\\
e^{i[\frac{k(r^2+\rho_{c,j}^2)}{2z}+\phi(\rho_{c,j})+(\mu_j-2)\frac\pi4]}
\label{eq:aaf:asy:006}
\end{multline}
where $\rho_{c,j}$ are the solutions of the equation $k\rho_c/z_f+\phi'(\rho_c)=0$, $z_{c,j}=-k/\phi''(\rho_{c,j})$ is the longitudinal position where the rays emitted from $\rho_{c,j}$ contribute to the caustic with $z_{c,1}<z_f<z_{c,2}$, and $\mu_j = \sgn(z_{c,j}-z)$. 
It is interesting to note that at the focus the Airy-type solution transforms into a superposition of two Bessel functions. 

\minititle{Nonparaxial autofocosuing}
Recently abruptly autofocusing beams have been studied also in the nonparaxial domain. 
It has been shown that, due to the large bending angles of the rays, such optical waves can focus at a smaller distance, and can have a smaller spot size and a larger intensity contrast, as compared to the paraxial case~\cite{penci-ol2016}. 

\minititle{Different classes}
By engineering the profile of radially symmetric beams on the input plane it is possible to generate autofocusing-type beams with different characteristics. In particular, abruptly autodefocusing beams whose intensity abruptly decreases by several orders of magnitude right after the focus were predicted~\cite{chrem-pra2012}. Selecting both the diverging and the converging rays on the input plane, optical bottle beams both in the paraxial~\cite{chrem-ol2011-fourier} and in the nonparaxial regime~\cite{penci-ol2018} were generated. Such micro-scale optical bottle beams are particularly useful for optical trapping and manipulation of single or small number of absorbing particles.

Interestingly, the concept of autofocusing beams is not limited to beams with an on-axis peak. Rather, by introducing a vortex phase to the autofocusing beams, the focus transforms into a vortex ring structure, as was shown theoretically~\cite{jiang-oe2012,chen-oe2015} and experimentally~\cite{davis-oe2012}. In a similar vein, various classes of autofocusing beams with power exponent helical phase have also been demonstrated~\cite{li-oe2014}. Radially and azimuthally polarized circular Airy beams with a high numerical aperture were investigated theoretically and experimentally~\cite{wang-apb2014}. It was found that, near the focal region, a long subwavelength dark channel with full width at half maximum about $0.49\lambda$ is generated with depth of focus about $52\lambda$.

In a related notion, higher order cusp catastrophes of the Pearcey type have been examined and found to exhibit autofocusing properties~\cite{ring-oe2012,chen-ol2018}. Likewise, abruptly autofocusing waves can also exhibit an abrupt polarization transition at the focal point, and the associated orbital angular momentum converts into spin angular momentum~\cite{liu-ol2013}. Such a transition takes place when the topological charges of the polarization and the attached spiral phase are equal in number.

\section{Nonlinear dynamics}\label{sec:nonlinear}

\minititle{Theory}
As the intensity of light increases nonlinear light matter interactions start to become significant. The Kerr nonlinear properties of accelerating waves were first studied in~\cite{giann-pla1989}. In that work, it was shown that the nonlinear Schr\"odinger equation in an accelerating frame of reference transforms to the Painlev\'e II transcendent $y''=xy\pm 2y^3$, which is a direct nonlinear generalization of the Airy differential equation. It was found that, in contrast to bright soliton solutions, such localized states exist for both self-focusing and self-defocusing nonlinearities. These nonlinear extensions of the Airy beam are also diffraction-free and carry infinite power.
However, this field blossomed only twenty years later~\cite{kamin-prl2011} with the surge of research activity on Airy beams, when it became natural to ask whether diffraction-free accelerating beams can also exist in nonlinear media. Early attempts, in theory and experiments, have shown that launching Airy beams into a nonlinear medium inevitably deform and cannot accelerate in  a shape-preserving form~\cite{hu-ol2010-pr,chen-pra2010}. Eventually, in 2011, accelerating self-trapped beams were predicted theoretically in Kerr, saturable, and quadratic media~\cite{kamin-prl2011}.
Specifically, it was shown that accelerating self-trapped beams propagate in a stable fashion in Kerr media under self-defocusing conditions. On the other hand, in the self-focusing regime, the dynamics depend on the strength of the nonlinearity. For moderate or weak nonlinearities such states exhibit stable propagation, whereas for strong nonlinearities they emit solitons while they continue to accelerate. This latter phenomenon of soliton emission was investigated in the case when the system is excited with an Airy beam or pulse~\cite{fatta-oe2011}. It was found that soliton shedding from an Airy beam occurs in a medium with a focusing nonlinearity, depending on the launching parameters. By varying the initial beam intensity, zero, one, or more than one solitons are emitted from the Airy wave which propagate towards the oscillatory part of the solution. In spite of the power loss, due to soliton shedding, the main lobe of the Airy wave is recovered during propagation due to self-healing. The propagation of a 2+1D Airy beam in a nonlinear medium was also considered~\cite{chen-pra2010} by using the method of moments and direct numerical simulations. The nonlinear dynamics of Airy beams~\cite{hu-ol2012} and pulses~\cite{hu-ol2013} have been analyzed experimentally by exploring the spectral properties of the wave during propagation. In the case of two transverse dimensions, the effect of wave collapse was examined in~\cite{chen-sr2013}. Airy waves in different types of nonlocal nonlinearities have also been studied~\cite{beken-oe2011,shen-sr2015}. 

\minititle{Experiments}
The dynamics of Airy beams in the presence of a saturable self-focusing and self-defocosuing photorefractive nonlinearity have been experimentally reported in~\cite{hu-ol2010-pr}. It was shown that in the case of a self-defocusing nonlinearity, the Airy beam can maintain its shape during propagation. 
The experimental observation of self-accelerating beams in quadratic media in 1+1D and 2+1D was presented in~\cite{dolev-prl2012}.
The quadratic nonlinearity is fundamentally local, hence it would be expected that, for second harmonic generation with a shaped beam, the intensity peaks of the second harmonic would follow those of the first harmonic. However, the prediction~\cite{kamin-prl2011} and the experimental observation~\cite{dolev-prl2012} of shape-preserving accelerating beams supported by the quadratic nonlinearity have shown otherwise: the intensity peaks of the first and second harmonics are asynchronous with respect to one another, yet the coupled harmonics dp  exhibit joint acceleration on the same curved trajectory within the quadratic medium. Furthermore, the self-healing properties of such vectorial structures is investigated. Nonlinear three-wave mixing process have also been employed in the generation of an Airy beam from a pump under phase matching conditions~\cite{ellen-np2009,dolev-apl2009}. 
During harmonic generation of abruptly autofocusing beams it was shown that the high harmonics preserve the phase distribution of the fundamental beam~\cite{koulo-prl2017}. 

\minititle{Losses}
The presence of nonlinear losses can reduce the intensity of the main lobe of the Airy beam during propagation. However, it was predicted that it is possible to compensate for nonlinear losses by appropriately selecting the truncation function leading to the generation of a paraxial stationary accelerating beam~\cite{lotti-pra2011}. Furthermore, even in the non-paraxial domain, it was demonstrated that it is possible to compensate for both linear and nonlinear losses for the generation of loss-proof accelerating beams~\cite{schle-nc2014}.

\minititle{Unbiased photorefractive crystals}
A different class of accelerating nonlinear closed-form solutions of the Airy type was proposed in unbiased photorefractive crystals~\cite{chris-ol1996}. The solutions exist due to the diffusion-induced photorefractive nonlinearity for beam intensities well above the dark irradiance. In~\cite{jia-prl2010} such accelerating Airy waves were experimentally observed in a strontium barium niobate crystal.

\minititle{Fibers: Raman + Cross phase modulation}
The use of Airy pulses in intrapulse Raman scattering control was examined in~\cite{hu-prl2015}. The possibility of tuning the frequency of a laser pulse via the use of an Airy pulse-seeded soliton self-frequency shift was demonstrated. When compared with the use of conventional chirped Gaussian pulses, this technique brings about unique advantages in terms of efficient frequency tuning, along with the generation and control of multicolor Raman solitons with enhanced tunability. 
In~\cite{gouts-jo2017} it was shown that the velocity and thus the frequency of a signal pulse can be adjusted by the use of a control Airy pulse. The whole process is controllable by using Airy pulses with different truncations that leads to predetermined values in the frequency shift. 

Recently, a self-accelerating optical pump wave-packet was employed to demonstrate controlled shaping of one type of generalized Cherenkov radiation~\cite{hu-sr2017}. By tuning the parameters of the wave-packet, the emitted waves can be judiciously compressed and focused at desired locations. The presence of higher-order effects in the dynamics of Airy pulses has been examined in~\cite{zhang-josab2014,zhang-oe2014-mi,zhang-oe2014-raman}.

\section{Non-uniform media}\label{sec:nonuniform}

It is of practical importance to analyze the dynamics of accelerating beams beyond the case of a uniform index distribution. 
In this section we focus on the propagation of accelerating beams on interfaces between two different media, in media with particular classes of refractive index distributions, and in media with a periodic index modulation. 

\minititle{Trajectory engineering - dynamics in potentials} 
The presence of a linear index potential modifies the trajectory of an Airy beam but does not really affect its diffraction-free characteristics. Specifically, it has been demonstrated that in the presence of a linear index potential the acceleration can be increased, reduced, canceled, or even reversed~\cite{liu-ol2011,ye-ol2011}. The more general problem of a linear index potential that dynamically changes along the propagation direction as $d(z)x$ was treated in~\cite{efrem-ol2011}. Even in this case, the Airy beam remains diffraction-free. In addition, the trajectory of the Airy beam can follow any arbitrary predefined path by appropriately engineering the functional form of $d(z)$. The manipulation of an Airy plasmon by a linear potential was demonstrated in~\cite{bleck-ol2013}. In the presence of a parabolic index the Airy wavepacket was shown to exhibit periodic inversion of its profile as well as periodic self-Fourier transformation~\cite{zhang-oe2015,zhang-ap2015}. The properties of abruptly autofucusing beams can also be controlled in the presence of a linear~\cite{hwang-ieee_pj2012} or a parabolic potential~\cite{zhong-oe2016}. The reflection and transmission dynamics of an Airy beam impinging on refractive index potentials was studied in~\cite{efrem-pra2014}. Two generic classes of potentials were studied: localized potentials whose contrast reduces to zero outside a specific region and sigmoid-type potentials. 
In general the parabolic trajectory of the Airy beam is not maintained by the reflected and transmitted waves with the exception of reflection from piecewise linear potentials. 

\minititle{Reflection from interfaces} 
The reflection and refraction of a finite power Airy beam on the interface between two dielectric media was investigated in~\cite{chrem-josaa2012}. Through numerical simulations, the self-accelerating dynamics of the Airy-like refracted and reflected beams were examined. In the case of a Brewster angle it was found that the reflected beam achieves self-healing, despite the severe suppression of a part of its spectrum. In the total internal reflection regime the beam remains nearly unaffected except for the Goos-H\"anchen shift. In the case of nonlinear interfaces, giant values of the Goos-H\"anchen shifts are displayed~\cite{chamo-ol2014}. These shifts have an extreme sensitivity to the input intensity and exhibit multiple critical values.

\minititle{Lattices}
Curved beams are also supported in photonic lattices. In~\cite{ganai-pra2011} it was shown that in the absence of a linear index gradient, the features of the Wannier-Stark modes mainly self-bend along two opposite hyperbolic trajectories. The accelerating properties of the Wannier-Stark states were observed in semiconductor-like photonic superlattices~\cite{qi-ol2014}. Accelerating beams with an engineered convex shape and with an aberration-free focus can be generated in a lattice by utilizing catastrophe optics calculations~\cite{efrem-ol2012}. Beyond the coupled mode theory approach the presence of higher order bands can affect the dynamics of the beams~\cite{chrem-pra2012}. Specifically, aberration-free focal spots can concentrate light from even higher bands, while a single caustic can originate from multiple bands. Beams with an accelerating profile have been found to exist in photonic crystals~\cite{kamin-oe2013}. Such waves retain their profile while bending to highly non-paraxial angles along a circular-like trajectory. Under the action of nonlinearity, the beam structure and acceleration cannot be preserved. In~\cite{makri-ol2014} it was shown that linear and nonlinear diffraction-free self-accelerating beams can exist in the case where both the lattice and the diffraction-free beam follow the same parabolic trajectory. 

\section{Applications}\label{sec:applications}

A main driving force promoting research in this area is the variety of proposed applications in different branches of optics. Below we discuss different classes of applications that make use of the properties of accelerating and abruptly autofocusing waves. 

\begin{figure}
\centering
\includegraphics[width=\columnwidth]{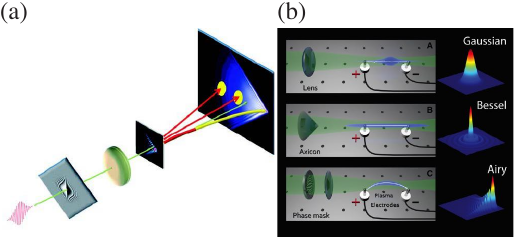}
\caption{(a) Generation of a curved plasma channel by an Airy type filament. The conical radiation emitted from the curved filament along the tangent of the channel trajectory can be spatially resolved on the observation plane. (b) Electric discharges following the plasma channels created by a focused Gaussian beam, a Bessel beam and an Airy beam. The Airy beam generates a curved plasma channel.
  Reprinted (a) with permission from Science~\cite{polyn-science2009} and (b) adapted from~\cite{cleri-sa2015}. 
}
\label{fig:filamentation}
\end{figure}
\subsection{Filamentation}\label{subsec:filamentation}

Filamentation occurs when an intense laser beam propagates in a dielectric medium with a sufficient intensity to cause ionization~\cite{couai-pr2007}. Optical filaments have a wide spectrum of applications ranging from remote sensing to THz generation and lighting control. In this respect the use of accelerating and autofocusing beams allows for a high degree of filamentation control. 

\minititle{Accelerating filament}
By utilizing a high intensity two-dimensional Airy beam in air, a plasma channel has been generated that follows a curved parabolic trajectory~\cite{polyn-science2009}. One of the important attributes of laser-induced filaments is the forward emission of broadband light, a property that enables various applications in remote spectroscopy. Analysis of the conical emission emanating from straight filaments is complicated by the fact that emissions originating from different longitudinal sections of the beam overlap in the observation plane. However, as can be seen in Fig.~\ref{fig:filamentation}(a), the conical radiation emitted from the curved filament along the tangent to the trajectory channels can be spatially resolved on the observation plane. The filamentation of Airy beams in water has also been reported in~\cite{polyn-prl2009}. Due to the Airy nature of the filament, the generated supercontinuum is mapped in the far field revealing the dynamical characteristics of the filament.

\minititle{Electric discharges}
The curved trajectory of an Airy beam has been used to guide electric discharges along complex curved trajectories~\cite{cleri-sa2015}. 
In manufacturing, arc discharges are used for machining, micromachining, and for assisting the milling process. In addition, such a precise trajectory control can be useful for delivering electric discharges to specific targets, and for electronic jamming. The use of diffraction-free Airy beam results into a more uniform trajectory for the discharge as compared to a Gaussian beam as shown in Fig.~\ref{fig:filamentation}(b). Such Airy-induced arcing can even circumvent an object that completely occludes the line of sight.

\minititle{Autofocusing filaments}
When focusing high intensity Gaussian beams, the focus position suffers from severe and uncontrolled shifting due to nonlinear effects. On the contrary, by using abruptly autofocusing waves, it has been found that the position of the focal spot exhibits only minor changes as a function of the input power~\cite{panag-nc2013}. After the focal spot, the autofocusing beam transforms to a quasi-uniform intense light bullet, in contrast to the case of Gaussian beams where the resulting wave exhibits severe spatiotemporal reshaping during propagation. Furthermore, such autofocusing ring-Airy wavepackets are able to overcome the reference intensity clamping of $4\times10^{13}$ W/cm$^2$ associated with filaments generated by Gaussian beams, at low numerical apertures and form an intense sharp on-axis focus~\cite{panag-pra2016}. 

\minititle{THz from autofocusing}
One of the main applications of optical filaments is the generation of plasma-induced THz radiation, normally by using Gaussian beam excitation. However, as it was shown in~\cite{liu-optica2016}, the use of a two-color abruptly autofocusing beam greatly enhances the THz wave pulse energy generated by the plasma in air by 5.3 times, leading to new possibilities for efficient THz generation and applications.

\begin{figure}
\centering
\includegraphics[width=\linewidth]{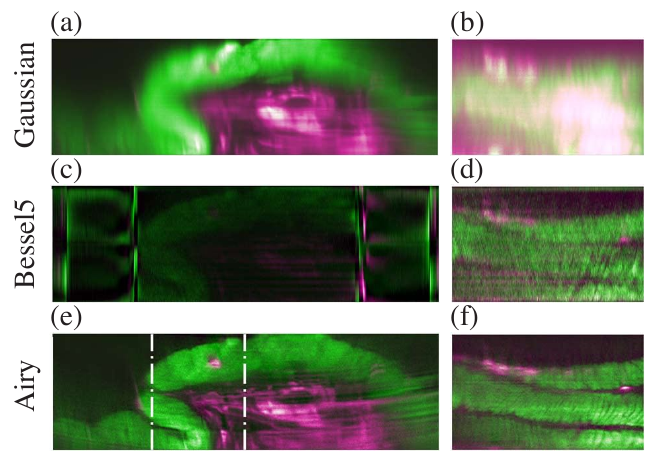}
\caption{Light sheet microscopy scan of a juvenile amphioxus. Comparison of the images taken by using (a)-(b) Gaussian, (c)-(d) Bessel, and (e)-(f) Airy illumination beams.
  Reprinted with permission from Springer Nature~\cite{vette-nm2014}.
}
\label{fig:imaging}
\end{figure}
\subsection{Imaging}\label{subsec:imaging}

\minititle{Extended depth of field}
In 1995, a method was proposed to extend the depth of field by using a cubic phase mask~\cite{dowsk-ao1995,cathe-ao2002}. It was shown that the resulting point-spread function and the optical transfer function do not change significantly as a function of misfocus. Although not explicitly mentioned, the point-spread function generated in the far field takes the form of a two-dimensional finite energy Airy beam, which is the reason why it does exhibit appreciable changes in the out-of-focus areas.

\minititle{Super-resolution fluorescence imaging}
Super-resolution fluorescence imaging techniques overcome the diffraction limit and allow fluorescence imaging with near molecular-scale resolution. Various strategies have been used to localize fluorescent molecules in three dimensions for 
super-resolution imaging, which however suffer from either anisotropic image resolution or a limited imaging depth. A method that utilizes an Airy beam point spread function was demonstrated in~\cite{jia-np2014}. The main reasoning for using Airy beams as point spread functions is that they can propagate over several Rayleigh lengths without appreciable diffraction and are self-healing after being obscured in scattering media. Using this method, super-resolution imaging with isotropic three-dimensional localization precision of $10-15$ nm over a $3$ $\mu$m imaging depth from $\sim2000$ photons per localization was obtained using biological samples~\cite{jia-np2014}. The Airy-type point spread function was found to provide a two- to three- fold higher $z$  localization precision than other conventional 3D localization approaches that use point spread function engineering. 

\minititle{Light-sheet microscopy}
Light-sheet microscopy facilitates high-contrast volumetric imaging with minimal sample exposure by using two orthogonal objectives that perform the task of excitation and detection. The factor that determines the axial resolution is the width of the illuminated light sheet. Usually Gaussian beams are used, which are limited by the Rayleigh range that severely restricts the uniformly illuminated field of view. Using diffraction-resisting Bessel beams one can create a uniformly thin light sheet but the transversal outer ring structure produces background fluorescence and, in principle, precludes high axial resolution. It was shown that utilizing an accelerating Airy beam provides a wide field of view in which the transverse beam structure contributes positively to the imaging process~\cite{vette-nm2014}. The comparison shown in Fig.~\ref{fig:imaging} demonstrates the clear advantage of Airy beams. An additional factor that limits the penetration of optical fields to the tissue is losses. As it was demonstrated in~\cite{nylk-sa2018}, by utilizing an Airy-type accelerating beam with a parabolic trajectory and attenuation-compensated amplitude 
profile~\cite{preci-ol2014}, 
it is possible to generate a uniform intensity envelope with an eightfold improvement of the contrast-to-noise ratio across the entire field of view in thick biological specimens. Note that currently, there is a commercially available Airy beam light sheet microscopy imaging system (Microscopy: Aurora).

\begin{figure}[t]
\centering
\includegraphics[width=\columnwidth]{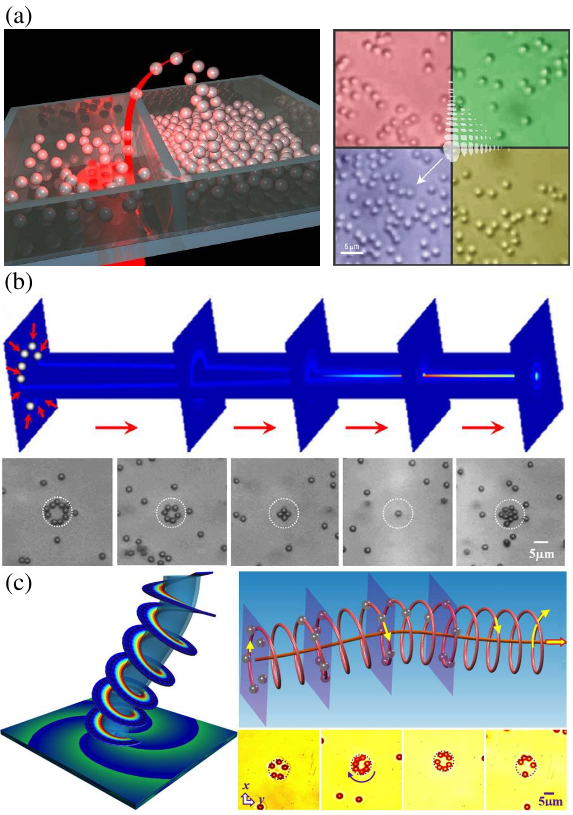}
\caption{(a) An Airy beam acting as a snowblower conveying particles from one compartment to another (left panel). Micrographs of the sample after colloids were exposed to the Airy beam, incident from below, for 2 min (right panel). (b) Experimental observation of particle guidance with a circular autofocusing beam. (c) A diffraction-resisting singular beam with a helical phase structure along the curve (illustrated in the left panel) used for 3D microparticle guiding and spiraling driven by the singular beam propagating along a hyperbolic secant trajectory (right panels).
}
\label{fig:particle}
\end{figure}

\subsection{Particle manipulation}\label{subsec:particle}

Perhaps, one of the most exciting proposed applications of Airy beams is particle transportation and manipulation, thanks to their self-bending properties and self-healing capabilities even in adverse environments such as turbulent and scattering media~\cite{broky-oe2008,hu-ol2010}. Demonstrations of particle manipulation based on Airy beams not only fueled further research interest in accelerating beams, but also added a new tool in the arena of optical trapping and manipulation, where light induced forces play a crucial role~\cite{grier-nature2003}.

\minititle{Particle cleaning}
The apparent acceleration of Airy wavapackets was utilized in~\cite{baumg-np2008} to optically mediate particle propulsion and targeted redistribution along parabolic trajectories via the combined action of the gradient force and radiation pressure. 
In the experiment, an Airy beam was used as a micro-scale ``snowblower'' to transport  particles along the beam path. As shown in Fig.~\ref{fig:particle}(a), an Airy beam, represented by the white overlay (its deflection during propagation is shown by the white arrow), was launched onto a monolayer of 3 $\mu$m silica colloids. Over a time scale of minutes, the particles in the upper right box (green) were pushed in both vertical (against gravity, into the page) and transverse directions, resulting in the upper right box emptying and the lower left (purple) box being filled~\cite{baumg-np2008}. The extent of such targeted particle clearing is of course significantly enhanced because of the non-diffracting nature of the Airy beams as opposed to Gaussian beams. In addition, due to the self-healing property of Airy beams, such optically mediated particle clearing could be carried out more robustly. In subsequent experiments, Airy beams were exploited for optical redistribution of microparticles and red blood cells between specially designed microwells~\cite{baumg-lc2009}. The ability of Airy beams in trapping particles along parabolic trajectories was quantitatively analyzed in later studies, including the use of a tightly focused Airy beam in an optical tweezers setting which resulted in an inversion of the particle distribution before and after the focal plane~\cite{cheng-oe2010,zheng-ao2011}.

\minititle{Autofocusing and vortex Bessel-like}
In addition to the conventional 2D Airy beams, other families of self-accelerating beams have also been employed for optical trapping and manipulation. 
In this respect, abruptly autofocused beams were used to guide and transport microparticles along the primary rings of a radially symmetric Airy beam~\cite{zhang-ol2011}. As illustrated in Fig.~\ref{fig:particle}(b), the particles are first trapped by the brightest ring, and then transported along the autofocusing beam. From the snapshots, it can be seen that at different longitudinal positions, as the radius of the circular ring decreases, less particles can stay trapped in the ring. At the ``focal point'', when the beam abruptly focuses into a bright spot, only one particle can stay in the trap. These results suggest that the abruptly autofocused Airy beams can be used as a tapered channel guide for microparticles. Furthermore, by Fourier-transforming an appropriately apodized Bessel beam, abruptly autofocusing beams and elegant fully-closed bottle beams can be created~\cite{chrem-ol2011-fourier}, which were employed as a robust photophoretical trap for airborne particles and to transport trapped particles from one glass container to another~\cite{zhang-oe2012}. 
Utilizing vortex Bessel-like beams, 3D manipulation of microparticles was experimentally demonstrated as shown in Fig.~\ref{fig:particle}(c). In this case the particles follow a spiraling trajectory along the main high intensity ring of the higher-order Bessel beam with a center that follows parabolic, hyperbolic and even three-dimensional trajectories~\cite{zhao-sr2015,zhao-sb2015}. 

\minititle{Nonparaxial accelerating}
It is worth mentioning that for many applications pertaining to optical trapping and manipulation, it is desirable to use nonparaxial accelerating beams to achieve large bending angles. In fact, nonparaxial particle manipulation in fluids with steeper angles was demonstrated using the so-called loss-proof self-accelerating beams~\cite{schle-nc2014}. 
In transparent media, these beams exhibit an exponential intensity growth, which may facilitate other novel applications in micromanipulation and ``ignition'' of nonlinear processes.

\begin{figure}
\centering
\includegraphics[width=0.6\columnwidth]{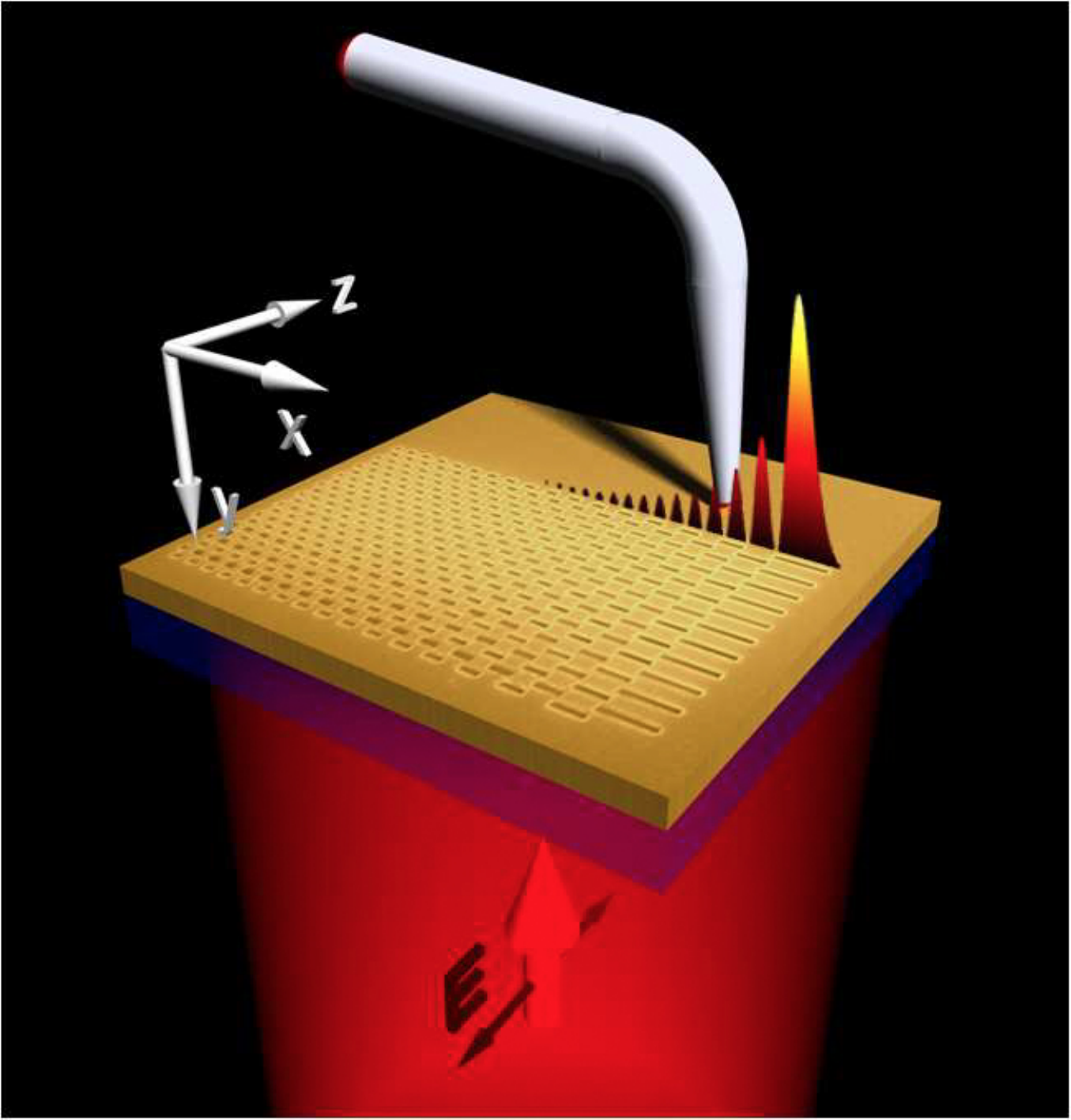}
\caption{Generation of Airy plasmons. Schematic of the experimental setup with the engineered grating.
  Reprinted with permission from American Physical Society~\cite{minov-prl2011}.
}
\label{fig:plasmons}
\end{figure}
\subsection{Plasmons}\label{subsec:plasmons}

\minititle{Proposal}
A surface plasmon polariton (SPP) refers to a strongly localized collective motion of free-electrons when coupled to photons at a metal-dielectric interface. 
Being excited with an Airy beam, the surface-plasmon polariton inherits its remarkable properties. Airy plasmons were first proposed in~\cite{salan-ol2010}, when it was realized that the strictly vectorial problem of surface plasmon propagation at a metal-dielectric planar interface could be formally reduced to a scalar one. As in the case of free space Airy beams, Airy plasmons represent the only plasmonic wave that is simultaneously non-spreading, self-bending, and self-healing. The latter property leads to a field configuration that is relatively unaffected by surface roughness and fabrication defects of the metallic layer. In addition, it has been shown that Airy plasmons can be excited over a broad set of beam sizes and wavelengths. 

\minititle{Observation}
Although it is a challenge to excite such Airy plasmons experimentally, as a wave vector mismatch between an SPP and a free space wave should be compensated by a suitable coupling technique, only one year later three research groups successfully demonstrated the observation of such Airy plasmons utilizing different excitation methods~\cite{zhang-ol2011-plasmon,minov-prl2011,li-prl2011}. 
In particular in~\cite{zhang-ol2011-plasmon} a free-space 1D Airy beam was directly coupled onto a metal surface through a grating coupler, and demonstrated that the SPPs can be guided along a metallic interface without using any physical waveguide structures. The Airy wave was impinged onto a gold grating with a period of $805$ nm, and then the excited Airy plasmons were directly monitored via leakage radiation microscopy. Importantly, it was shown that the ballistic trajectory 
as well as the peak intensity position of the Airy plasmons can be reconfigured in real-time. 
In~\cite{minov-prl2011} a different grating coupling scheme was utilized (see Fig.~\ref{fig:plasmons}), in which a grating pattern was designed is such as way so as to imprint the phase profile of an Airy function, thus exploiting the uniqueness and self-healing properties of the Airy plasmons. The grating was excited from the substrate side with an extended beam at a wavelength of $784$ nm, and the intensity distribution was probed with a near-field scanning optical microscope with single photon counting capabilities. 
Finally, in~\cite{li-prl2011} the Airy plasmon was excited by using a carefully designed nanoarray structure (an array of nanocavities with chirped separation), which allowed for in-plane propagation of Airy-like SPP waves along the air-silver interface~\cite{li-prl2011}. In addition, a new phase-tuning method based on a non-perfectly matched diffraction process that is capable of generating and modulating plasmonic Airy beams at will was proposed. 

\minititle{Other works}
Thus far, considerable research efforts have been devoted to accelerating plasmonic Airy waves. These include, for instance, the generation and manipulation of Airy plasmon trajectories in a linear potential or a gradient system~\cite{liu-ol2011,bleck-ol2013}, and the interference of two Airy plasmons which led to the generation and control of a bright plasmonic hot spot on the surface of a metal film~\cite{klein-ol2012}. In addition to the normal configuration of excitation of Airy plasmons, a variety of other techniques has been proposed and demonstrated, including arbitrarily shaped surface plasmons based on holographic beam shaping~\cite{dolev-prl2012-plasmon}, highly localized Cosine-Gauss Plasmon Beams~\cite{lin-prl2012}, collimated plasmon beams by means of in-plane diffraction with symmetric phase modulation~\cite{li-prl2013}, and bending plasmonic beams along arbitrary paraxial and nonparaxial curvatures~\cite{epste-prl2014,libst-prl2014}. Some of these works have been previously discussed in~\cite{minov-lpr2014}. Different methods for generating accelerating light with plasmonic metasurfaces have been suggested and implemented. Accelerating beams at the second harmonic were also produced using nonlinear metasurfaces based on split-ring resonators~\cite{keren-acs2016}. In addition accelerating light inside glass was generated by using a metasurface consisting of plasmonic nanoantennas~\cite{henst-optica2018}.

\begin{figure}
\centering
\includegraphics[width=\columnwidth]{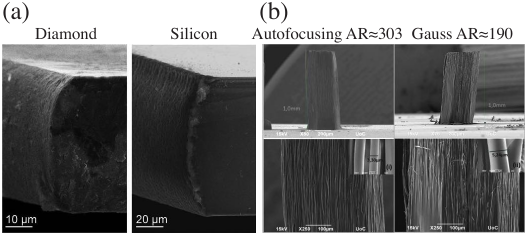}
\caption{Material processing. (a) Machining of diamond and silicon using circular accelerating beams. For diamond the radius of the circular profile is 70 $\mu$m and the sample thickness 50 $\mu$m and for silicon the radius is 120 $\mu$m and the sample thickness is 100 $\mu$m. (b) Use of abruptly autofocusing beams for multi-scale photo-polymerization. For comparison of the fabricated structures are shown using autofocusing (left) and Gaussian beams (right). Compared to the Gaussian beam the Airy beam presents an almost twice as large focal voxel aspect ratio (AR) while it results in much better fabrication quality.
  (a) Reprinted with permission from AIP Publishing~\cite{mathi-apl2012},
  (b) reprinted with permission from OSA~\cite{manous-optica2016}. 
}
\label{fig:material}
\end{figure}

\subsection{Material processing}\label{subsec:material}

\minititle{Accelerating beams for machining}
The properties of accelerating and abruptly autofocusing beams have been proven very useful in various applications pertaining to material processing with laser beams. These include for example, ablation, machining, and fabrication. 
The use of high power lasers for material ablation is widely applied in a broad range of technologies in material processing. A particular challenge is the machining of structures both on the surface and within the sample that have longitudinally varying characteristics because this requires simultaneous control of beam steering and sample rotation and translation. A solution to this problem that relies on the use of accelerating beams was demonstrated in~\cite{mathi-apl2012}. Specifically, diamond and silicon samples were machined to a curved profile as can be seen in Fig.~\ref{fig:material}(a). In addition, accelerating beams have been applied for the writing of curved trenches to $\sim80$ $\mu$m depth within the bulk sample.

\minititle{Autofocusing beams for ablation and fabrication}
The use of abruptly autofocusing waves for creating ablation spots on the back side of thick samples was demonstrated in~\cite{papaz-ol2011}. It was shown that the low intensity profile before the focus, the long working distance, combined with the short focal volume and the high focal intensity, makes abruptly autofocusing waves ideal candidates for material processing.  
By changing the parameters of the Airy ring it is possible to generate abruptly autofocusing beams with a tunable focus and practically the same focal shape and volume. Such optical waves have been utilized for the fabrication of large three-dimensional structures with high resolution using multi-photon polymerization~\cite{manous-optica2016}. It was shown that, as compared to Gaussian and Bessel beams, the autofocusing beam presents a superior focal control, with high aspect ratio, with respect to positioning and shape [Fig.~\ref{fig:material}(b)]. In addition, such beams do not require very large numerical aperture that limits the height of the generated structure.

\minititle{Airy pulses for nanochannels}
The key to initiate material processing is the deposition of a sufficient energy density within the electronic system. In dielectrics this critical energy corresponds typically to a plasma frequency in the near-IR spectral region. The use of Airy pulses has the advantage that the main lobe is constantly refilled with energy as compared to regular Gaussian pulses that have limited penetration. Airy pulses were utilized in~\cite{gotte-optica2016} to efficiently generate high aspect ratio nanochannels. Depths in the range of several micrometers and diameters around $250$ nm are created with a single laser shot without making use of self-focusing and filamentation processes. 

\subsection{Other applications}\label{subsec:other}

\minititle{Supercontinuum generation} 
The supercontinuum generation with femtosecond Airy pulses in highly nonlinear microstructured optical fibers was demonstrated in~\cite{ament-prl2011}. The self-healing ability of the Airy pulse was shown to play an important role in the process. The pulse dynamics were found to be different for the cases when the tail of the waveform propagates in front of or behind its main peak. Specifically, when the tail precedes the main peak several soliton emissions take place from the main Airy pulse. The cross-phase modulation interaction of the solitons leads to the appearance of blue shifted spectral peaks from the emitted dispersive wave. On the other hand, if the tail trails the main peak then the soliton generated from the first collapse of the main peak passes through the tail of the waveform and disrupts its ability to self-regenerate.

\minititle{Airy beam laser}
A design for generating a laser that produces two-dimensional Airy beams was proposed and demonstrated in~\cite{porat-ol2011}. The principle relies on the use of a reflection diffraction grading as the laser output mirror. A binary mask is designed to add a cubic phase to the diffraction light. A lens that performs a Fourier transform generates the Airy beam at the output.  

\minititle{Routers-interconnects} 
Using an optical induction technique, it was demonstrated that Airy beams can be utilized as all-optical routers with individually addressable output channels~\cite{rose-apl2013}. In addition, by activating multiple channels at the same time, an optically induced splitter is created with configurable outputs. Along similar lines, in~\cite{wiers-ol2014} the use of two incoherent counter-propagating Airy beams as optically induced waveguide structures was analyzed for all-optical interconnects. 

\section{Accelerating waves in different settings}\label{sec:settings}
\minititle{Electron wavepackets} 
Besides optics, accelerating waves have been suggested and observed in different physical settings. The first prediction of an Airy wave packet was given in the context of quantum mechanics~\cite{berry-ajp1979}. Interestingly, if the wavefunction of a free quantum particle has the form of an Airy function, its shape is preserved while it freely accelerates along a parabolic trajectory. Such quantum wavepackets have been observed using free electrons~\cite{voloc-nature2013}. In particular, electron Airy beams were generated by diffraction of electrons through a nanoscale hologram, which imprinted on the electrons' wavefunction a cubic phase modulation in the transverse plane.

\minititle{Dirac equation}
The dynamics of self-accelerating Dirac particles was studied in~\cite{kamin-np2015}. It was shown that engineering the wavefunction of electrons, as accelerating shape-invariant solutions to the potential-free Dirac equation, this fundamentally acts as a force. As a result, the electrons accumulate an Aharonov-Bohm-type phase-which is equivalent to a change in the proper time and is related to the twin-paradox gedanken experiment. This implies that fundamental relativistic effects such as length contraction and time dilation can be engineered by properly tailoring the initial conditions. 

\minititle{Acoustic waves}
Acoustic self-bending along arbitrary convex trajectories and bottle waves have been reported in~\cite{zhang-nc2014}. For bottle beams, the sound energy flows through a three-dimensional curved shell in air leaving a close-to-zero pressure region in the middle, thus circumventing an obstacle. Accelerating sound waves have also been demonstrated in~\cite{zhao-sr2014}. The setup consists of an array of loudspeakers which are modeled as point sources. Acoustic paraxial Airy beams as well as nonparaxial waves with circular trajectories have been observed. On a related topic, accelerating Airy beams were also observed with underwater acoustic waves~\cite{bar-prb2015}. 

\minititle{Linear and nonlinear water waves}
Linear and nonlinear surface gravity water waves having the form of an Airy envelope were observed in~\cite{fu-prl2015}. In the linear regime, the shape of the envelope is preserved while propagating in an $18$ m water tank. In the nonlinear regime, the position-dependent chirp and cubic phase shift are directly observed in these systems. In this case, the dynamics are governed by a nonlinear Schr\"odinger-type equation, better known as the Dysthe equation.

\minititle{Curved space}
Accelerating beams have also been predicted to exist in curved space~\cite{beken-prx2014}. Exact solutions were found that propagate along accelerating (or more precisely non-geodesic) trajectories. Such wave packets exist due to the interplay between space curvature and interference. Recently such beams have been experimentally observed on spherical surfaces~\cite{patsy-prx2018}.

\minititle{Bose-Einstein condensates}
It had also been predicted that Bose-Einstein condensates can support self-accelerating matter waves without the presence of any external potential~\cite{efrem-pra2013}. Different classes of accelerating and abruptly autofocusing waves were proposed and different schemes for the preparation of the condensate using laser beams (to imprint an amplitude or a phase pattern onto the matter wave) were suggested. The effect of mean-field interactions was examined and it was found that, independent of the type and strength of the nonlinearity, the dynamics was associated with the generation of accelerating matter waves. In the case of strong attractive interactions, the acceleration was found to increase while the radiation can reorganize itself in the form of solitons.

\section{Conclusions}
In the last ten years or so, the field of curved and accelerating waves has been the focus of intense investigation. Since the first prediction and experimental demonstration of Airy beams in optics, this field has experienced a remarkable growth, starting from new theoretical developments in producing specifically designed classes of optical accelerating waves, to different generation methods and experiments. More importantly, accelerating waves are nowadays finding applications in many and diverse areas. These include for example, filamentation, imaging and microscopy, particle manipulation, plasmonics, material processing, and supercontinuum generation, to mention a few.
Equally important, the concept of shape-preserving accelerating wavepackets has been extended to very many kinds of wave systems in nature, ranging from plasma waves to sounds waves, and from gravity waves in liquids to a plethora of quantum waves systems. In many of these, the underlying dynamics is very different than the original context where Airy beams were discovered.
In this article, we provided a brief overview of these recent developments, while many new fundamental findings and applications are still to be discovered in this exciting field of accelerating waves.

\section*{Funding Information}
Z.C. is supported by the National Key R\&D Program of China (Grants No. 2017YFA0303800). The work of D.N.C. and M.S. has been partially supported by the US-Israel Binational Science Foundation (Award \# 2016381).

\newcommand{\noopsort[1]}{} \newcommand{\singleletter}[1]{#1}

\end{document}